\documentclass[a4paper,11pt]{article}
\usepackage{jinstpub} % for details on the use of the package, please
\usepackage{lineno}
\title{\boldmath Observation of the cosmic ray shadow of the Sun with the ANTARES neutrino telescope}

\author[a,b]{A. Romanov}
\author[a,b]{and M. Sanguineti}

\affiliation[a]{Dipartimento di Fisica dell’Università,\\Via Dodecaneso 33, 16146 Genova, Italy}
\affiliation[b]{INFN - Sezione di Genova,\\Via Dodecaneso 33, 16146 Genova, Italy}

\emailAdd{andrey.romanov@ge.infn.it, matteo.sanguineti@ge.infn.it}

\abstract{ANTARES is the largest undersea neutrino telescope and it has been taking data in its final configuration for more than ten years. On their journey to the Earth, cosmic rays can be absorbed by celestial objects, like the Sun, leading to a deficit in the atmospheric muon flux measured by the ANTARES detector, the so-called Sun "shadow" effect. This phenomenon can be used to evaluate fundamental telescope characteristics: the detector angular resolution and pointing accuracy. This work describes the study of the Sun "shadow" effect using the ANTARES data collected between 2008 and 2017. The statistical significance of the Sun shadow observation is $3.7\sigma$ and the estimated angular resolution value of the ANTARES telescope for downward-going muons is  $0.59^{\circ} \pm 0.10^{\circ}$, which is consistent with the expectations obtained from the Monte Carlo simulations and also with the estimation from the Moon "shadow" analysis of 2007-2016 years. No evidence of systematic pointing shift is found and the resulting pointing accuracy is consistent with the expectations.}

\collaboration[c]{on behalf of the ANTARES Collaboration}

\proceeding{Very Large Volume Neutrino Telescopes Workshop}

\begin{document}
\maketitle
\flushbottom

\notoc

\section{Introduction}
\label{sec:intro}

Detection of the neutrino point-like sources is one of the main goals of the ANTARES neutrino telescope \cite{Main}. To achieve this aim, it is highly important to have a reliable way to estimate the angular resolution of the telescope. This work describes the study of the Sun shadow effect. The term shadow stands for the deficit in the secondary atmospheric muon flux in the direction of the Sun. The deficit is caused by the absorption of the primary cosmic rays by the celestial body. By measuring this deficit, an appropriate estimation of the telescope angular resolution can be obtained \cite{ice, HAWC, argo}.  This work presents the Sun shadow analysis using the ANTARES 2008-2017 data sample, corresponding to a total detector livetime of 2925 days. The analysis is performed using one- and two-dimensional approaches, based on $2.6\times10^6$ events reconstructed as downward-going muons \cite{Sun_shadow}.

\section{The Sun shadow analysis}
\label{sec:main}

%\section{The data selection optimisation}
%\label{sec:mc}

The quality of each reconstructed track is described by means of two parameters: the likelihood-wise parameter, $\Lambda$, and the angular error estimator of the reconstructed direction, $\beta$ \cite{quality}. In order to find which set of cut values on $\Lambda$ and $\beta$ maximises the sensitivity to the Sun shadow detection, a Monte Carlo (MC) simulation is used. The simulation is done with the MUPAGE code based on parametric formulas \cite{mupage}. 

%The data-taking periods of the ANTARES detector are subdivided into samples lasting for several hours, \textit{runs}. The simulation follows the environmental conditions and the detector configuration, the so-called run-by-run simulation \cite{RbR}. In order to be able to perform the MC simulation during the reasonable CPU time, statistics of the MC sample is reduced to 1/3 of the actual run statistics. However, in this work the MC statistics is artificialy enlarged using the \textit{additional zones} approach. Original MC sample contains the atmospheric muons from the sky regions around the nominal Sun position. One can exploit the additional regions of the sky shifting the Sun position in time, with a step of 2 hours. Therefore, the whole statistics of MC considering 11 additional zones together with the Sun zone is 4 times larger than the real data statistics.

The quality cut optimisation is based on the hypothesis test approach. The null hypothesis $H_0$ corresponds to the background only (no-Sun hypothesis), while the $H_1$ hypothesis implies the presence of the Sun shadow. Two MC samples are generated corresponding to the two hypotheses. According to the $H_1$ hypothesis, the shadowing is introduced into the simulated sample by removing all the muons generated within the Sun disk (radius $0.26^\circ$). The test statistic is calculated under the assumption that the event population in each bin asymptotically follows a Gaussian probability distribution. It is defined under two hypotheses, $H_0$ and $H_1$, mentioned above, resulting in $\lambda_{0}$ and $\lambda_{1}$.

\begin{figure}[htbp]
\centering
\includegraphics[width=0.4\textwidth]{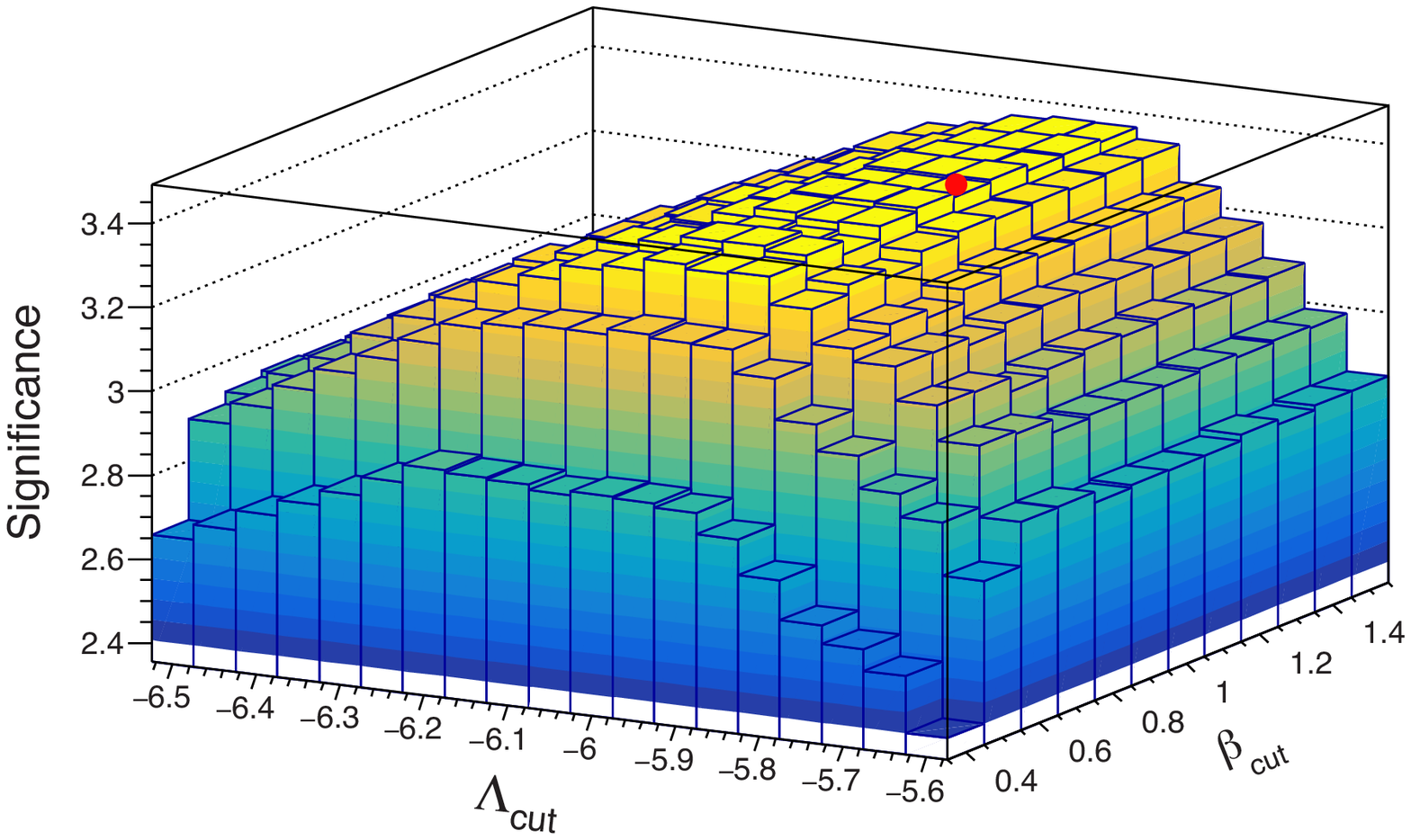}
\includegraphics[width=0.4\textwidth]{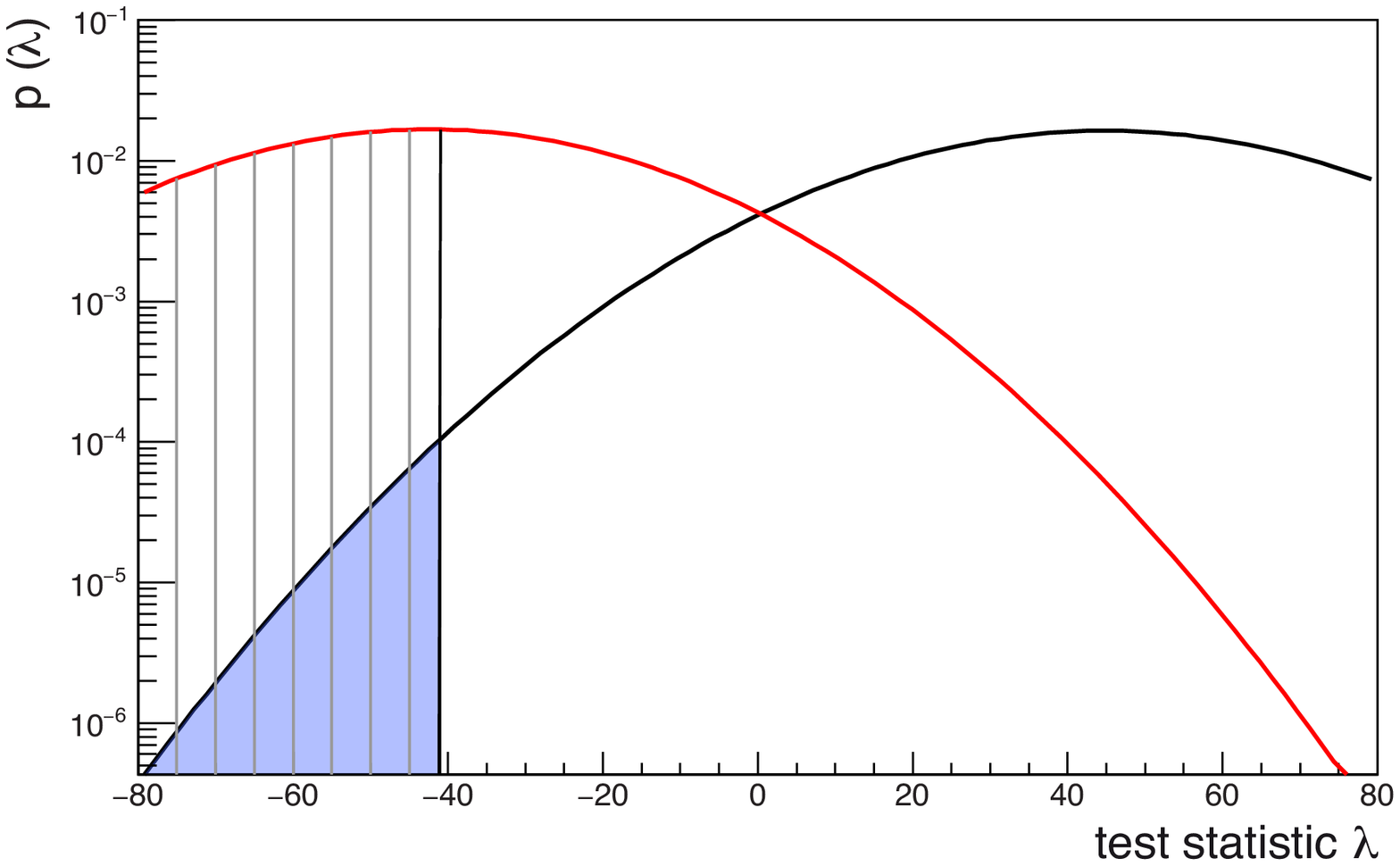}
\caption{\label{fig:cuts_and_lambda} Left: Expected statistical significance of the Sun shadow detection as a function of cut values on $\Lambda$ and $\beta$ ($\Lambda_{\rm{cut}}$ and $\beta_{\rm{cut}}$). The red point represents the selected set of cut values ($\Lambda_{\rm{cut}}=-5.9$ and $\beta_{\rm{cut}}=1.1^{\circ}$). Right: Distribution of the test statistic $\lambda$ for the two hypotheses, $H_0$ (black curve) and $H_1$ (red curve), obtained for the optimized set of cut values. The dashed area represents the fraction of pseudo-experiments ($50\%$) where $H_1$ hypothesis is correctly identified. The coloured area corresponds to the expected median significance ($3.4\sigma$) to reject the $H_0$ hypothesis in favour of the $H_1$ hypothesis.}
\end{figure}

This procedure is repeated assuming different set of quality cut values resulting in the distributions of $\lambda_0$ and $\lambda_1$. The set of cut values that maximises the expected significance is $\Lambda_{\rm{cut}}=-5.9$ and $\beta_{\rm{cut}}=1.1^{\circ}$, with the significance value of $3.4\sigma$ (Figure \ref{fig:cuts_and_lambda}, left). The corresponding distribution of $\lambda_0$ and $\lambda_1$ is shown in the right side of Figure \ref{fig:cuts_and_lambda}.

%This procedure is repeated assuming different set of quality cut values resulting in the distributions of $\lambda_0$ and $\lambda_1$. The expected statistical significance of the Sun shadow detection is calculated through the computation of the p-value of the $\lambda_0$ distribution corresponding to the median of the $\lambda_1$ distribution, for which $50\%$ of the pseudo-experiments under the $H_1$ hypothesis (presence of the Sun shadow) are correctly identified. The set of cut values which maximises the expected significance is $\Lambda_{\rm{cut}}=-5.9$ and $\beta_{\rm{cut}}=1.1^{\circ}$, with the significance value of $3.4\sigma$ (Figure \ref{fig:cuts_and_lambda}, left). The corresponding distribution of $\lambda_0$ and $\lambda_1$ is shown in the right side of Figure \ref{fig:cuts_and_lambda}.

\begin{figure}[htbp]
\centering 
\includegraphics[width=0.5\textwidth]{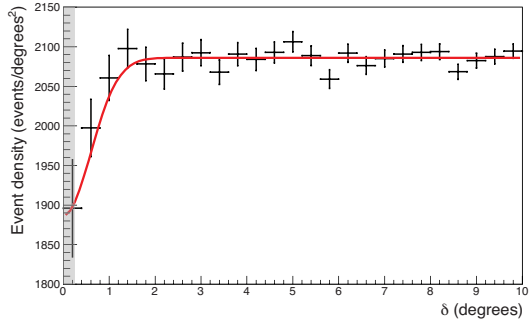}
\caption{\label{fig:fit} The muon event density as a function of the angular distance $\delta$ from the Sun centre based on the data sample taken in period 2008-2017 fitted with Eq. \ref{eq:gaussian1} (red line). The shaded area corresponds to the Sun angular radius ($0.26^\circ$).}
\end{figure}

The angular resolution of the ANTARES telescope for downward-going muons is estimated using the 2008-2017 data sample. The reconstructed events are selected with the optimised quality cuts described above, providing $6.5\times10^5$ events. The data event density distribution is produced in the same way as for the MC events described above in the hypothesis test procedure (Figure \ref{fig:fit}). 

The distribution is fitted with the following function, assuming the average geometrical Sun radius, $R_{\rm{Sun}}=0.26^\circ$: \cite{Moon_Antares}:

\begin{equation}
f(\delta)=k(1-\frac{R^2_{\rm{Sun}}}{2\sigma_{\rm{res}}^2}e^{-\frac{\delta^2}{2\sigma_{\rm{res}}^2}}), 
\label{eq:gaussian1}
\end{equation}

\noindent where $k = 2086 \pm 3$ is the average muon event density in the region far from the nominal Sun position and $\sigma_{\rm{res}}$ is the angular resolution value for downward-going muons. 

The value of $\sigma_{\rm{res}}$ from the fit is $0.59^\circ\pm0.10^\circ$, compatible with the one obtained in the ANTARES Moon shadow analysis ($0.73^\circ\pm0.14^\circ$) \cite{Moon_Antares}. The goodness of the fit is $\chi^2/\rm{dof}=19.6/23$. 

%In order to estimate the effect of a finite-size Sun radius pseudo-experiments were performed. The result of pseudo-experiments shows that the discrepancies between the assumed detector angular resolutions and the fitted values of the Gaussian width are below $10\%$ for the assumed angular resolution values above $0.35^{\circ}$, i.e. negligible with respect to the statistical uncertainty. Therefore, the obtained value of $\sigma_{\rm{res}}$ can be treated as the angular resolution of the telescope for downward-going muons. 

%The angular resolution value $\sigma_{\rm{res}}$ obtained in this work is compatible with the one obtained in the ANTARES Moon shadow analysis ($0.73^\circ\pm0.14^\circ$) \cite{Moon_Antares}.

The statistical significance of the result is estimated using the hypothesis test approach. For the $H_0$ hypothesis no shadowing effect is assumed. The significance is calculated from the test statistic function: $-\lambda = \chi^2_0 - \chi^2_1$, where $\chi^2_1$ is the value corresponding to the $H_1$ obtained from the fit with Eq. \ref{eq:gaussian1} and $\chi^2_0$ is the value corresponding to the horizontal line fit. The test statistic follows a $\chi^2$ distribution with 1 degree of freedom and the corresponding significance is $3.7\sigma$.

The Sun magnetic field could lead to the smearing of the shadow \cite{Ice2021}. In order to estimate such effect, the ANTARES data is subdivided into two samples covering the low (2008-2011) and the high (2011-2015) Sun activity. However, the statistics in these samples is insufficient to obtain significative conclusions.

%\section{Absolute pointing}
%\label{sec:2d}

\begin{figure}[htbp]
\centering
\includegraphics[width=0.4\textwidth]{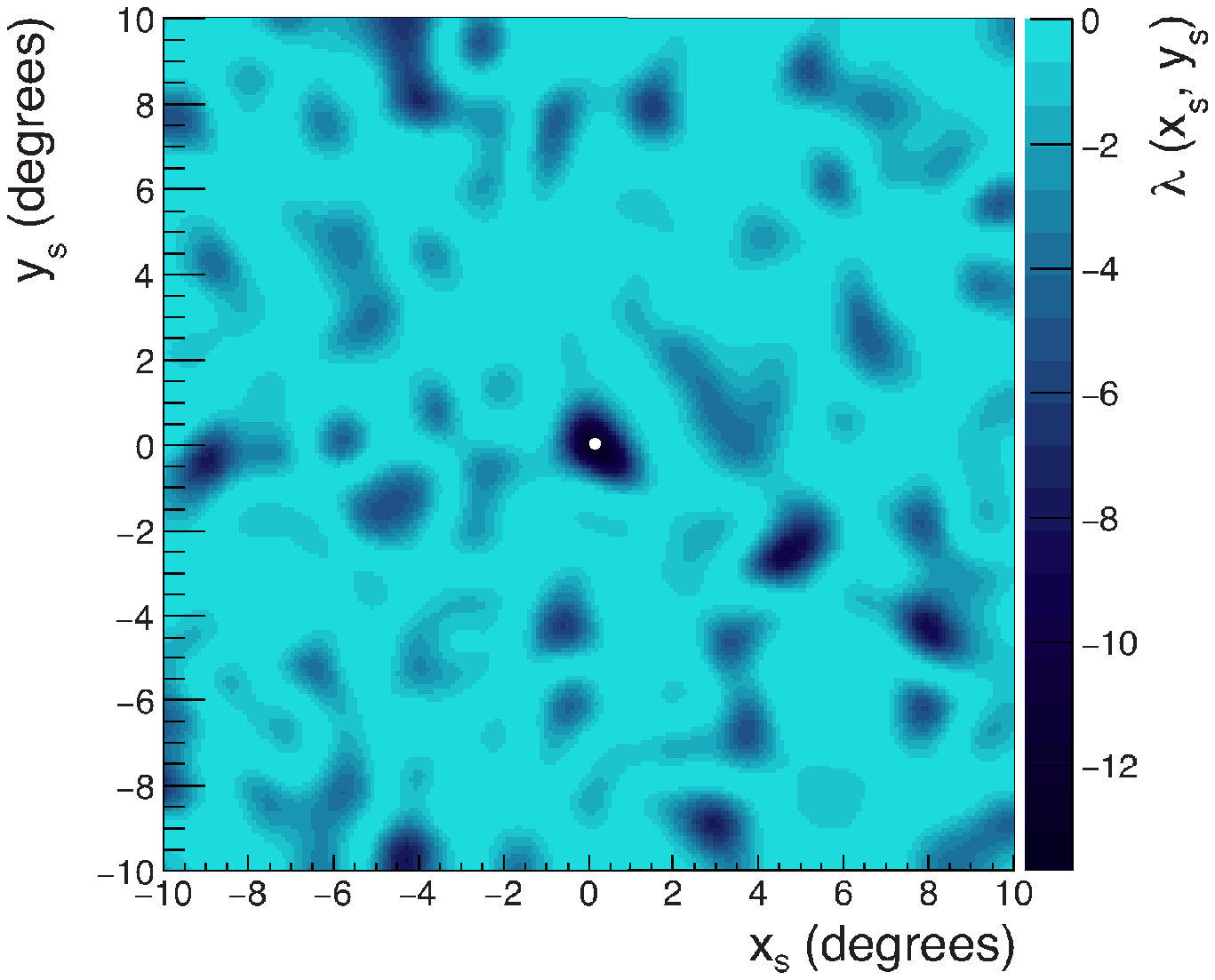}
\includegraphics[width=0.4\textwidth]{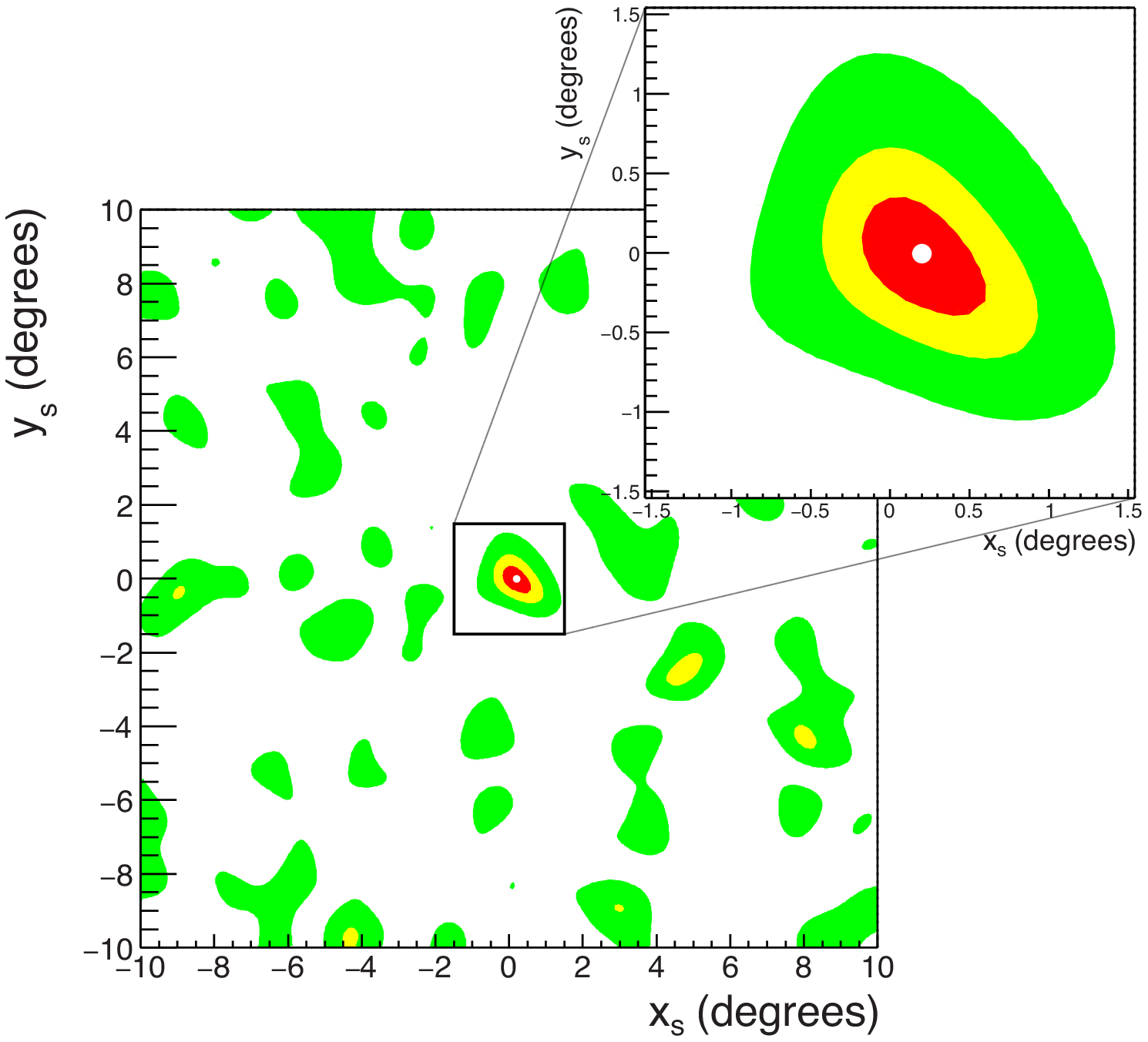}
\caption{\label{fig:two-d} Left: The distribution of the test statistic values around the nominal Sun position $O\equiv(0^\circ,0^\circ)$. The minimum value $\lambda_{\rm{min}}=-13.7$ is found at $(0.2^\circ,0^\circ)$ point (white dot). Right: Contours corresponding to different confidence levels (red: $68.27\%$; yellow: $95.45\%$; green: $99.73\%$). The white dot indicates $(0.2^\circ,0^\circ)$ point for which a minimum value of $\lambda_{\rm{min}}=-13.7$ is obtained.}
\end{figure}

The measurement of the Sun shadow allows also to estimate the detector pointing performance using a two-dimensional approach. For this purpose, the event distribution is projected on a two-dimensional histogram. The Sun shadow centre is assumed to be on each point of the histogram with a step size of $0.1^\circ$.  The nominal Sun position is $O\equiv(0^\circ,0^\circ)$ point.
The test statistic function, $\lambda(x_s,y_s)$, is then calculated for each assumed shift of the Sun position as the difference between $\chi^2$ values obtained from the fit under the $H_0$ and $H_1$ hypotheses. The minimum value of $\lambda(x_s,y_s)$ is found at $(0.2^\circ,0^\circ)$ point (Figure \ref{fig:two-d}, left).  

%At each bin, $-\lambda$ follows the distribution of a $\chi^2$ with one degree of freedom. This allows the estimation of the significance to reject the no-Sun hypothesis. Considering $-\lambda_{O}$, a p-value of $3.1\times10^{-4}$ is obtained. The corresponding significance is $3.6\sigma$.

The distribution of values of the test statistic $\lambda(x_s,y_s)$ can be interpreted as a bi-dimensional profile-likelihood. Therefore, the interval corresponding to a desired confidence level (CL) is obtained for $\lambda(x_s,y_s)$ (Figure \ref{fig:two-d}, right).

\section{Conclusions}
\label{sec:2d}

The Sun shadow effect is studied by means of two complementary approaches, one- and two-dimensional, which allow to determine the angular resolution for downward-going atmospheric muons and to verify the pointing performance of the detector. The effect is observed in the ANTARES 2008-2017 data with the statistical significance of $3.7\sigma$. The angular resolution for downward-going muons is $0.59^\circ\pm0.10^\circ$, compatible with the one found in the ANTARES Moon shadow analysis ($0.73^\circ\pm0.14^\circ$). No evidence of systematic pointing shift is found and the resulting pointing accuracy is consistent with the expectations.

\end{document}